\documentclass[twocolumn,amsmath,nosuperscriptaddress,amssymb,nopacs,nofootinbib,prl]{revtex4-2}

\usepackage[usenames,dvips]{color}
\usepackage{graphicx}
\usepackage{dcolumn}
\usepackage{bm}
\usepackage{mathtools}
\usepackage{epstopdf}
\usepackage{esint}
\usepackage{xcolor}
\usepackage{dsfont}
\usepackage{nicefrac}

\usepackage{amsmath,amssymb,amsfonts,mathrsfs}
\usepackage{ulem}
\usepackage{braket}
\usepackage{bbm}

\usepackage[colorlinks=true,citecolor=magenta,linkcolor=magenta, urlcolor=magenta]{hyperref}

\newcommand{\bea}{\begin{eqnarray}}
\newcommand{\eea}{\end{eqnarray}}
\newcommand{\bt}{\textbf}

\newcommand{\noi}{\noindent}
\newcommand{\no}{\nonumber}

\begin{document}
\def\v#1{{\bf #1}}

\title{Survival and Detection of Symmetry-Protected Topology in Loop Quenches}

\author{Nicol\`o Forcellini}
\email{nforcellini@baqis.ac.cn}
\author{Mikl\'os Horv\'ath}
\email{mikloshorvath@baqis.ac.cn}
\author{Panagiotis Kotetes}
\email{kotetes@baqis.ac.cn}

\affiliation{Beijing Academy of Quantum Information Sciences, Beijing 100193, China}

\vskip 1cm

\begin{abstract}
We explore a class of dynamical protocols - that we term \textit{loop quenches} - which are tailored for the study of symmetry-protected topological (SPT) systems. In loop quenches, SPT phases can survive even out of equilibrium, thus evading the dynamical violation of their protecting symmetry. Moreover, we demonstrate that employing loop quenches allows to detect the equilibrium topology via measurable dynamical quantities. Focusing on chiral-SPT phases, we introduce the Loschmidt chirality amplitude as a key observable that encodes the equilibrium topological invariant. We exemplify our method for chiral-symmetric one-dimensional two-band insulators and propose a pump-probe measurement scheme which allows to extract the amplitude in question. Our protocol uncovers a direct dynamical signature of SPT phases and, most importantly, paves the way for a general diagnostic framework that can be extended to other symmetry classes and dimensions.
\end{abstract}

\maketitle

\textit{\bt{Introduction} -} The non-trivial properties of pro\-to\-ty\-pi\-cal topological insulators~\cite{Hasan2010,Qi2011,KongRev}, such as, HgTe quantum wells, ${\rm Bi_2Se_3}$, and ${\rm FeTeSe}$ compounds, crucially rely on the presence of time-reversal symmetry (TRS). Preser\-ving a symmetry is also pivotal for the non-trivial to\-po\-lo\-gy of the celebrated Su-Schrieffer-Heeger (SSH) model. However, instead of an anti-unitary TRS, here, a unitary chiral symmetry (CS) is involved~\cite{AsbothBook}. All the above systems belong to the broader category of symmetry-protected to\-po\-lo\-gi\-cal (SPT) phases~\cite{Senthil2015,Wen2017}. These harbor frac\-tio\-na\-li\-zed boundary modes, which are topologically-protected against weak di\-sor\-der, as long as the symmetry of the bulk remains intact. The topological pro\-per\-ties of equilibrium SPT phases can be comprehensively understood and predicted using the ten-fold classification program~\cite{Ryu2010,Teo2010,KitaevClassi,Fidkowski}, which accounts for the possible CS and anti-unitary symmetries that are present in the system.

More recently, the study of non-equilibrium SPT phases has drawn significant attention. Its purpose is to investigate whether SPT phases can survive when time-dependent perturbations are added. This is a na\-tu\-ral question to ask, since topological phases protected by an anti-unitary symmetry, such as TRS, are expected to be drastically affected by the arrow of time and the unavoi\-dable dynamical breaking of the symmetry. In fact, the same holds for systems with CS. The simplest way to study SPT phases out-of-equilibrium, is to subject them to a quantum quench. Several
prior works have discussed ways to infer the to\-po\-lo\-gy of a system using quantum quenches~\cite{Vajna2014,DeLuca2014,Vajna2015,Nakagawa2016,Wang2017,Sun2018,Yang2018,Hu2020,Zhang2020,Rossi2022quenchedSSH,Qiu2024}. Others have focused on addressing the robustness of SPT phases~\cite{McGinley2018,Gong2018,Sedlmayr2018,McGinley2019,Pastori2020,Ghosh2023,Lane2024}, thus, also resulting in a classification for out-of-equilibrium topology~\cite{McGinley2018,McGinley2019}. From the latter, one concludes that out-of-equilibrium the ten symmetry classes collapse to only three, i.e., those which at most preserve a charge-conjugation symmetry.

\begin{figure}[t!]
\begin{center}
\includegraphics[width=\columnwidth]{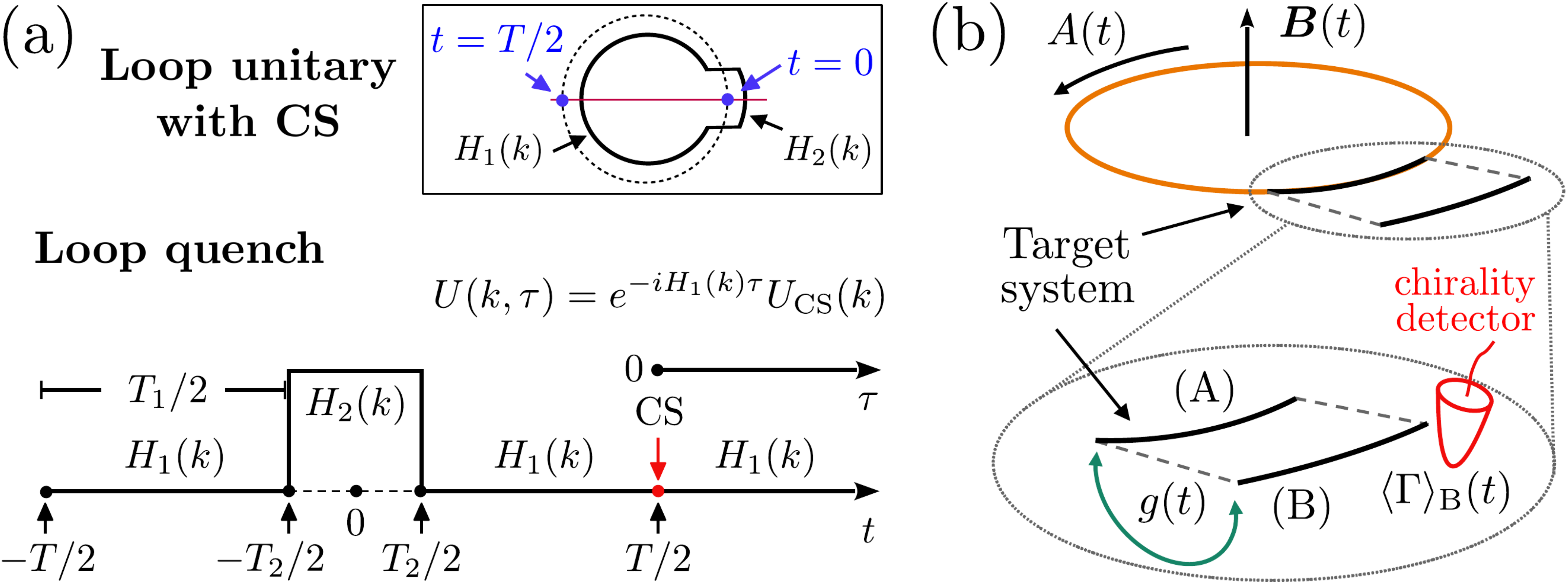}
\end{center}
\caption{(a) Time-profile for a loop quench, which is implemented by switching from the target system's Hamiltonian $H_1$, to the Hamiltonian $H_2$ of a pulse with a duration $T_2$, and then returning to $H_1$. Chiral symmetry (CS) is restored at $t=T/2$ and the system sees the evolution operator $U_{\rm CS}$. The inset depicts a loop unitary evolution with CS which is defined using the loop quench shown above. (b) The target system ${\rm (A)}$ is a one-dimensional electric-charge-conserving gapped SPT phase with CS. For its detection, we consider a ring geometry. A loop quench is effected by a time-varying out-of-plane magnetic field $\bm{B}(t)$. The target system (A) is probed by tunnel-coupling it to system (B). (A) and (B) are identical, but (B) is unaffected by the quench. To infer the topology of (A) one detects the induced chirality $\langle\Gamma\rangle$ on (B).}
\label{fig:Figure1}
\end{figure}

In this Letter, we bring forward a loophole to the above out-of-equilibrium classification. As we show, symmetry-protected topology can survive and can be detected even out-of-equilibrium by subjec\-ting the system to a special type of quench protocols, which we here refer to as \textit{loop quenches} (LQs). A LQ defines a dynamical process in which the Hamiltonian of the system of interest $H_1$, is subject to a ``pulse'' dictated by the Hamiltonian $H_2$, and subsequently returns to $H_1$. See Fig.~\ref{fig:Figure1}(a) for an illustration. Crucially, in contrast to the single-step $H_1\mapsto H_2$ quenches stu\-died previously, the dynamical violation of TRS and CS can be evaded in LQs. This becomes possible when the center of the pulse coincides with the center of the LQ, as in the inset of Fig.~\ref{fig:Figure1}(a). Such symme\-tric LQs are analogous to symmetric pe\-rio\-dic drives, which allow for dy\-na\-mi\-cal symmetries and Floquet SPT phases~\cite{Asboth2013,Asboth2014,Nathan2015,Fruchart2016,RoyHarper1,RoyHarper2,Yao2017,Kennes2019,Assili2024,Cardoso2025}. Most remar\-kably, by suitably designing the pulse Hamiltonian $H_2$, LQs offer an additional and far more promising functio\-na\-li\-ty. That is, to provide a quantized  measurable quantity which encodes the to\-po\-lo\-gi\-cal invariant of $H_1$, even when TRS and CS are dynamically broken. This holds in Fig.~\ref{fig:Figure1}(a) when the evolution goes beyond the TRS and CS restoration time.

In the following, we employ a concrete two-band model for an electric-charge-conserving one-dimensional insulator with CS, to exemplify the dynamical recovery
of the symmetry and the detection of the associated SPT phase. In addition, by assuming that the target system is expe\-ri\-men\-tal\-ly configurable in a ring geometry, we introduce a generic LQ-probe protocol that can be used to infer
the to\-po\-lo\-gi\-cal in\-va\-riant of its SPT phase in equilibrium. By residing on electric-charge conservation, we propose a LQ whose pulse consists of switching on a magnetic field. As depicted in Fig.~\ref{fig:Figure1}(b), here the field is orien\-ted out of the plane defined by the ring, so to ge\-ne\-ra\-te a vector potential in the system of interest. The topological properties of the insulator can be experimentally detected via quantum state transfer. This is achie\-va\-ble by weakly tunnel-coupling the target system to an identical copy of it, which, however, is unaffected by the LQ. Notably, the here proposed LQ-probe stra\-te\-gy can be generalized and extended to other types of SPT phases.

\textit{\bt{Loop quenches and chiral symmetry} -} We now proceed by de\-fi\-ning the LQs. The time-dependent Hamiltonian $H(k,t)$ possesses a dynamical CS generated by $\Gamma$, when the following relation is satisfied~\cite{Asboth2013,Asboth2014,Fruchart2016}:
\begin{align}
\Gamma^\dag H(k,t_R+t)\Gamma=-H(k,t_R-t),
\end{align}

\noi for all time instants $t$, where $\Gamma^2=\mathds{1}$ and consequently $\Gamma^\dag=\Gamma$. In the above, $t_R$ indicates some reference time. Without loss of generality, we set $t_R=0$ in the remainder. Consider now the cor\-responding time-evo\-lu\-tion ope\-ra\-tor (with the reduced Planck constant $\hbar=1$):
\begin{align}
U(k,t_i,t_f)={\cal T}\,{\rm Exp}\left[-i\int_{t_i}^{t_f}dt'\,H(k,t')\right],
\end{align}

\noi where ${\cal T}$ denotes time ordering. CS is dynamically preserved when the initial and final times $t_{i,f}$ are chosen symmetrically around $t=t_R=0$. That is, setting $t_i=-T/2$ and $t_f=+T/2$. For such a protocol, the time evolution can be divided into two segments with evolution ope\-ra\-tors $U(k,-T/2,0)$ and $U(k,0,+T/2)$. Note that these satisfy $U(k,0,+T/2)\equiv\Gamma U^\dag(k,-T/2,0)\Gamma$. As a result, one finds that $U(k,-T/2,+T/2)\equiv U_{\rm CS}(k)$ respects CS~\cite{Cardoso2025}, and fulfills $\Gamma U_{\rm CS}(k)\Gamma=U_{\rm CS}^\dag(k)$.

A simple illustration of the realization of a time evolution that preserves CS is given in Fig.~\ref{fig:Figure1}(a). There, we depict a chiral-symmetric LQ, where the initial Hamiltonian $H_1(k)$ first becomes quenched to $H_2(k)$ at $t=-T_2/2$ and subsequently ``loops back'' to $H_1(k)$ at $t=+T_2/2$. Since the Hamiltonian satisfies $H(k,-T/2)=H(k,T/2)$, it can be expanded in terms of a Fourier series in the interval $t\in[-T/2,T/2]$. This property allows to view the LQ protocol as a single-period evolution of a perio\-di\-cal\-ly driven system for which the Floquet theorem applies~\cite{OkaRev,RudnerRev}. Si\-mi\-lar\-ly to one-dimensional Floquet insulators with CS, here we can also  define a ``stroboscopic'' band invariant for the LQ which is obtained from the effective Hamiltonian $H_{\rm CS}(k)$ of $U_{\rm CS}(k)$. This is defined as $H_{\rm CS}(k)=\frac{i}{T}\ln\big[U_{\rm CS}(k)\big]$ and satisfies $\{H_{\rm CS}(k),\Gamma\}=0$.

However, truly periodically-driven insulators which respect CS support two band invariants instead. This is by virtue of the freedom to choose a chiral-symmetric frame, which stems from the time periodicity and allows us to relate the two band invariants to the quasi-energy invariants, with the help of which we can predict the to\-po\-lo\-gi\-cal zero and $\pi$ modes~\cite{Asboth2013,Asboth2014,Nathan2015,Fruchart2016,RoyHarper1,RoyHarper2,Yao2017,Kennes2019,Assili2024,Cardoso2025}. Here, in order to be in a position to evaluate the quasi-energy invariants, it is required to consider the complementary LQ, i.e., the one that is initialized by the pulse Hamiltonian $H_2(k)$, then subsequently switched to $H_1(k)$, before looping back to $H_2(k)$. Besides inverting the sequence of the Hamiltonians, $T_2$ and $T_1=T-T_2$ should also exchange roles. By combining the two complementary LQs one essentially compactifies the time evolution into a loop -- see inset of Fig.~\ref{fig:Figure1}(a) -- thus enabling the definition of a \textit{loop time-evolution operator} in analogy to the stroboscopic time-evolution operator of Floquet systems~\cite{RoyHarper1,RoyHarper2}.

While chiral-symmetric LQs can be understood by suitably extending results for topological driven systems, in the remainder, we are primarily interested in the more general time evolution, whose evolution operator is eva\-lua\-ted at a time $t=T/2+\tau$ greater than the CS-restoration time $t=T/2$. Thus, for $\tau\geq0$ we have:
\begin{align}
U(k,\tau)\equiv{\cal T}\,{\rm Exp}{\left[-i\int_{-T/2}^{+T/2+\tau}dt'\, H(k,t')\right]}.
\end{align}

\noi For LQs of the type shown in Fig.~\ref{fig:Figure1}(a), we find the compact form $U(k,\tau)={\rm Exp}\big[-iH_1(k)\tau\big]U_{\rm CS}(k)$. Our main next goal is to identify a suitable LQ, that will enable us to infer the properties of $H_1(k)$ with the help of $U(k,\tau)$.

\textit{\bt{Two-band example} -} At this stage, we exemplify the emergence of CS in LQs and we look for
a link between $H_1(k)$ and $U(k,\tau)$ by exploring a two-band model which is described by a Hamiltonian of the form $H(k,t)=\bm{d}(k,t)\cdot\bm{\sigma}$, where $\bm{\sigma}=(\sigma_x,\sigma_y,\sigma_z)$ is a vector of Pauli matrices. By identifying the CS operator $\Gamma$ with $\sigma_z$, we have $\bm{d}(k,t)=(d_{x}(k,t),d_{y}(k,t),0)$ at all times. For a LQ as in Fig~\ref{fig:Figure1}(a), the evolution operator becomes:
\begin{align}
U(k,\tau)=e^{-iH_1(k)(T_1/2+\tau)}e^{-iH_2(k)T_2} e^{-iH_1(k)T_1/2}\,,
\label{eq:EvoTwoBand}
\end{align}

\noi One can easily verify that the chiral-symmetry-breaking part of $U(k,\tau)$, i.e., the coefficient of the CS matrix $\sigma_z$, is given by the expression:
\begin{align}
\sin\big[d_1(k)\tau\big]\sin\big[d_2(k)T_2\big]\left[\hat{\bm{d}}_1(k)\times \hat{\bm{d}}_2(k)\right]\cdot\hat{\bm{z}},
\end{align}

\noi with $d_{1,2}(k)\equiv|\bm{d}_{1,2}(k)|$ and $\tau\geq0$, as shown in Fig.~\ref{fig:Figure1}(a). CS is recovered \textit{exactly} at $\tau=0$ because --  there -- the factor $\sin{[d_1(k)\tau]}$ becomes zero independently of the form of $\bm{d}_2(k)$. We remark that in the general case, where $d_1(k)$ is not
a constant function of $k$, there is no time instant other than $\tau=0$ at which CS is restored for all $k$.

As this point, it is particularly useful to obtain the matrix elements of $U(k,\tau)$ in the basis of the eigenstates $\big|\psi^\pm(k)\big>$ of $H_1(k)$ with eigenenergies $\pm d_1(k)$.
Straightforward manipulations lead to the result:
\bea
&&U^{\alpha\beta}(k,\tau)\equiv \big<\psi^\alpha(k)\big|U(k,\tau)\big|\psi^\beta(k)\big>\no\\
&&=e^{-id_1(k)[(\alpha+\beta)T_1/2+\alpha\tau]}\big<\psi^\alpha(k)\big|e^{-iH_2(k)T_2}\big|\psi^\beta(k)\big>,\,\quad
\eea

\noi where $\alpha,\beta=\pm$. In particular,
for the choice $\alpha=-\beta=+$ we find that the dependence with respect to $T_1$ drops out completely and the exponential prefactor simply becomes ${\rm Exp}[-id_1(k)\tau]$. Even more, we find the expression:
\bea
\big<\psi^+(k)\big|e^{-iH_2(k)T_2}\big|\psi^-(k)\big>\qquad\qquad\qquad\qquad\quad\no\\=\sin\big[d_2(k)T_2\big]\left[\hat{\bm{d}}_1(k)\times\hat{\bm{d}}_2(k)\right]\cdot\hat{\bm{z}}\,,
\label{eq:EvoCross}
\eea

\noi therefore also quantifying the dynamical violation of CS.

\textit{\bt{Loschmidt chirality amplitude and topology}} - The above results motivate us to introduce the here-termed Loschmidt chirality amplitude (LCA):
\begin{align}
{\cal G}_{\rm CS}(k,\tau)=\big<\psi^-(k)\big|\Gamma\big|\psi^-(k,\tau)\big>\equiv\big<\psi^+(k)\big|\psi^-(k,\tau)\big>\,,\label{eq:LCA}
\end{align}

\noi where $\big|\psi^\pm(k,\tau)\big>=U(k,\tau)\big|\psi^\pm(k)\big>$ constitute the time-evolved counterparts of $\big|\psi^\pm(k)\big>$. We remark that the equi\-va\-len\-ce in Eq.~\eqref{eq:LCA} is by virtue of CS which imposes $\Gamma\big|\psi^\pm(k)\big>=\big|\psi^\mp(k)\big>$ and, in turn, also implies ${\cal G}_{\rm CS}(k,\tau)\equiv U^{+-}(k,\tau)$. The LCA constitutes the complementary to the Loschmidt amplitude for systems with a CS, since, instead of relating the initial state to its time-evolved, it relates it to its time-evolved CS-partner.

From the result in Eq.~\eqref{eq:EvoCross}, we observe that the LCA is proportional to the cross product $\hat{\bm{d}}_1(k)\times\hat{\bm{d}}_2(k)$ which quantifies the violation of CS and the chirality production. At the same time, it is crucial to recall that there exists another cross product involving $\hat{\bm{d}}_1(k)$. This appears in the winding number $\nu_1$, which classifies $H_1(k)$:
\begin{align}
\nu_1=\int_{-\pi}^{+\pi}\frac{dk}{2\pi} \  \Big[\hat{\bm{d}}_1(k)\times \partial_k\hat{\bm{d}}_1(k)\Big]_z\,.
\end{align}

\noi Hence, in order to link ${\cal G}_{\rm CS}(k,\tau)$ to $\nu_1$, it is required to also link
$\bm{d}_2(k)$ to $\partial_k\bm{d}_1(k)$. This motivates
us to devise the following protocol. Consider a LQ which contains a magnetic-field pulse with duration $T_2$ which, in turn, induces the uniform vector potential $A$ in the ring:
\bea
H_2(k)=\bm{d}_1(k+eA)\cdot\bm{\sigma}\approx\big[\bm{d}_1(k)+eA\partial_k\bm{d}_1(k)\big]\cdot\bm{\sigma},\,\,\,
\eea

\noi where $e>0$ denotes the elementary unit of electric charge. The above approximation is justified for sufficiently weak strengths of $A$, which are much smaller than $E_g/e\upsilon_F$, where $\upsilon_F$ is the Fermi velocity and $E_g={\rm min}[d_1(k)]$ is the minimum of the band gap. We depict such a situation in Fig.~\ref{fig:Figure1}(b), where magnetic flux is threaded through the ring-shaped one-dimensional system of interest. The radius of the ring is substantially large to ensure that curvature effects are negligible and that the vector potential is uniform throughout it.

Within the above approximation, we find that at lowest order in $A$, the LCA can be written in the following form:
\begin{align}
\frac{{\cal G}_{\rm CS}(k,\tau)}{eAT_2}
\approx\frac{\sin\big[d_1(k)T_2\big]}{T_2}e^{-id_1(k)\tau}\Big[\hat{\bm{d}}_1(k)\times \partial_k\hat{\bm{d}}_1(k)\Big]_z.
\end{align}

\noi Inspired by the structure of the above expression, we introduce the one-sided retarded Fourier transform of the LCA in the frequency domain:
\begin{align}
{\widetilde{\cal G}_{\rm CS}}(k,\omega)=\int_0^\infty d\tau\, e^{i(\omega+i\eta)\tau}\,{\cal G}_{\rm CS}(k,\tau)\,,
\end{align}

\noi with $\eta$ being a regularization parameter, which is ideally a positive infinitesimal, that is, $\eta=0^+$. In practice, it suffices that $\eta$ is much smaller than any energy scale in the system. Of particular interest is the imaginary part of ${\widetilde{\cal G}_{\rm CS}}(k,\omega)$ at zero frequency, which takes the form:
\begin{align}
\frac{{\rm Im}\,{\widetilde{\cal G}_{\rm CS}}(k,\omega=0)}{eAT_2}=-\frac{\sin\big[d_1(k)T_2\big]}{d_1(k)T_2}\Big[\hat{\bm{d}}_1(k)\times \partial_k\hat{\bm{d}}_1(k)\Big]_z.
\end{align}

\noi Since, $\lim_{x\rightarrow0}\sin x/x=1$, we observe that by further considering a sufficiently short pulse, so that $T_2\ll\hbar/E_g$,
the r.h.s. appearing above becomes proportional solely to the winding number density. Hence, in this limit, the $k$-space integrated LCA per length becomes proportional to $\nu_1$, and leads to the relation:
\begin{align}
-\int_{-\pi}^{+\pi}\frac{dk}{2\pi}\left.\frac{{\rm Im}\,{\widetilde{\cal G}_{\rm CS}}(k,\omega=0)}{AT_2}\right|_{T_2\rightarrow0}=\frac{e}{\hbar}\,\nu_1\,,
\end{align}

\noi where we reintroduced the reduced Planck constant. The above result reveals how the topology of the target system can be encoded in the LCA obtained in a LQ.

\textit{\bt{Loop-quench-probe protocol} -} Having established a connection between the LCA and the winding number of the target system, it is natural to ask whether it is possible to turn this correspondence into a guide for performing an actual measurement protocol. To answer this important question, we observe that ${\cal G}_{\rm CS}(k,\tau)$ contains the overlap between the quenched and unquenched eigenstates of $H_1(k)$. Hence, one can alternatively see this matrix element as an overlap between states of two identical systems (A) and (B), which are both initially described by $H_1(k)$, but only one -- say (A) -- is affected by the LQ, as shown in Fig.~\ref{fig:Figure1}(b) for the case of a magnetic pulse.

Our proposed experimental strategy is to first prepare system (A) in the ground state $\big|\psi^-(k)\big>$ and subsequently quench it. Then, for $\tau\geq0$, the post-quench state of (A) is imprinted in (B), by switching on a weak tunnel-coupling at $\tau=0$, which is described by the Hamiltonian:
\begin{align}
H_{\rm AB}(k,\tau)=g(\tau)\big|\psi_{\rm A}(k)\big>\big<\psi_{\rm B}(k)\big|+g^*(\tau)\big|\psi_{\rm B}(k)\big>\big<\psi_{\rm A}(k)\big|.
\end{align}

\noi Due to the above coupling, system (B) will evolve into a superposition of its initial state, that is also chosen to be $\big|\psi^-(k)\big>$, and the state of the target system (A). To infer the topological properties of (A), one can experimentally measure the chirality of system (B), i.e., the expectation value of the CS operator $\langle\Gamma\rangle_{\rm B}(k,\tau)=\big<\psi_{\rm B}(k,\tau)\big|\Gamma\big|\psi_{\rm B}(k,\tau)\big>$, whose time derivative satisfies:
\bea
\partial_\tau\langle{\Gamma}\rangle_{\rm B}(k,\tau)&=&i\big<\psi_{\rm B}(k,\tau)\big|H_1(k)\Gamma\big|\psi_{\rm B}(k,\tau)\big>\no\\
&-&ig^*(\tau)\big<\psi_{\rm B}(k,\tau)\big|\Gamma\big|\psi_{\rm A}(k,\tau)\big>+{\rm h.c.}\,\quad\,
\label{eq:EvoGamma}
\eea

\noi For a sufficiently weak tunnel-coupling strength $|g(\tau)|$, we can consider that $\big|\psi_{{\rm A},{\rm B}}(k,\tau)\big>$ are obtained by discar\-ding the mixing of (A) and (B) in Eq.~(\ref{eq:EvoGamma}). Under this condition, we have $\big|\psi_{\rm B}(k,\tau)\big>={\rm Exp}[-iH_1(k)\tau]\big|\psi^-(k)\big>$ and $\big|\psi_{\rm A}(k,\tau)\big>=U(k,\tau)\big|\psi^-(k)\big>$, where $U(k,\tau)$ is given by Eq.~\eqref{eq:EvoTwoBand}. Hence, at lowest order in the tunnel-coupling strength $|g(\tau)|$ we find that the first contribution to $\partial_\tau\langle{\Gamma}\rangle_{\rm B}(k,\tau)$ drops out, and the chirality production in (B) only originates from its coupling to (A). By further assuming that after switching on the tunnel-coupling the strength is given by  the real profile $g(\tau)\equiv V_{\rm AB}e^{-\eta\tau}/2\pi$, we are now in a position to integrate Eq.~\eqref{eq:EvoGamma} in the interval $\tau\in[0,\infty)$, and find that the total chirality per length $\delta\langle{\Gamma}\rangle_{\rm B}$ which is produced in this time window, reads as:
\begin{align}
\delta\langle{\Gamma}\rangle_{\rm B}=\frac{V_{\rm AB}}{2\pi}\int_{-\pi}^{+\pi}\frac{dk}{2\pi}\,{\rm Im}\,\widetilde{\cal G}_{\rm CS}(k,\omega=0).\label{eq:chProd}\end{align}

From the above expression we observe that the produced chirality is proportional to the desired imaginary part of the zero-frequency LCA. Therefore, such a cross-system measurement allows one to infer $\nu_1$. Given the above result, we are motivated to introduce the rate of the chirality-electric current cross-system susceptibility:\begin{align}
{\cal M}=\frac{1}{T_2}\frac{\delta\langle{\Gamma}\rangle_{\rm B}}{\delta A}\,,
\end{align}

\begin{figure}[t!]
\centering
\includegraphics[width=\columnwidth]{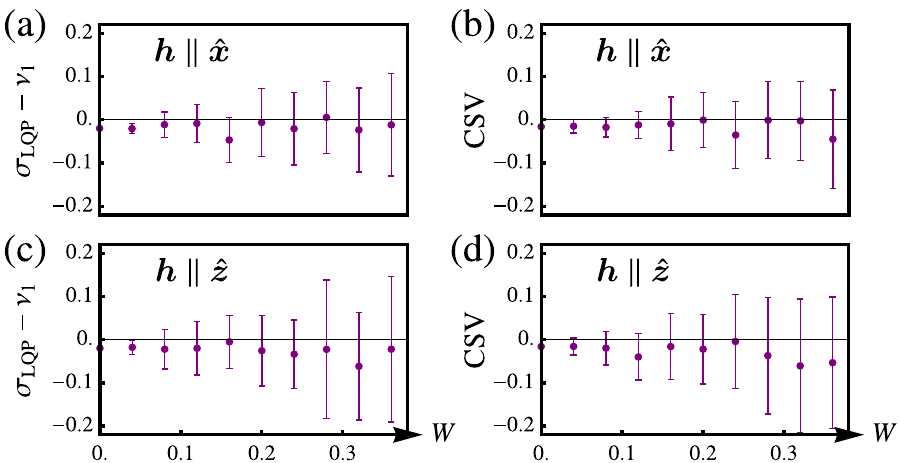}
\caption{Panels (a) and (c) show the deviation $\sigma_{\rm LQP}-\nu_1$, for a loop-quench-probe protocol which is initialized with $\bm{d}_1=(\sin{k}+0.1,-\cos{k},0)$, and enlists the Hamiltonian $H_1(k)$ in the AIII symmetry class. Here, $\sigma_{\rm LQP}$ is expressed in units of conductance $e^2/h$ and is averaged over $N=25$ realizations of the time-evolution with a random noise field $\bm{h}(t)$. Panels \{(a),(c)\} correspond to $\bm{h}(t)=w(t)\{\hat{\bm{x}},\hat{\bm{z}}\}$, with a noise strength $w(t)\in[-W,W]$. The results show stability around $\sigma_{\rm LQP}=1$ for noise strengths smaller than the minimum energy gap of the system, i.e., $W\ll1$. The error bars indicate the standard deviation obtained after averaging for the $N$ realizations. (b) and (d) show the outcome for the CSV measure, which quantifies the degree of CS violation during the time-evolution. The results for $\bm{h}||\hat{\bm{z}}$ appear less robust, as this is reflected in the larger standard deviation in (c) and (d) compared to (a) and (b). This is attributed to the fact that, for $\bm{h}||\hat{\bm{z}}$, CS is not only violated due to the time-profile, but also explicitly due to the orientation of $\bm{h}$. For the numerics we used $T_2=0.1$, $\eta =0.015$, and a cutoff time $T_{c}=300$.
}
\label{fig:Figure2}
\end{figure}

\noi which is expressed in the same units as the orbital magnetization~\cite{NiuRev}. Hence, for a sufficiently weak and brief pulse, the here-introduced LQ-probe conductance:
\begin{align}
\sigma_{\rm LQP}=-e\left(\frac{d{\cal M}}{dV_{\rm AB}}\right)_{A,T_2,V_{\rm AB}\rightarrow0}=\frac{e^2}{h}\,\nu_1\,,\label{eq:LQCond}
\end{align}

\noi becomes quantized in terms of the topological invariant of $H_1(k)$, thus, establishing the here-sought-after link between the LQ-probe protocol and the topology of (A). Note, that we intentionally chose the form of $g(\tau)$ as above and defined $\sigma_{\rm LQP}$ in such a manner, so to draw a parallel to the quantization of the transverse conductance arising in the quantum Hall effect. Here, the electrostatic energy $V_{\rm AB}$ which controls the tunneling ${\rm A}\leftrightarrow{\rm B}$ plays the role of the chemical potential in the quantum Hall effect.

\textit{\bt{Numerical simulations of a noisy loop-quench-probe protocol} - } We now explore the effects of noise on the inference of $\nu_1$ for a concrete two-band model. We incorporate noise by coupling the system to an external field $\bm{h}(t)=w(t)\bm{h}$, so that $H_{\rm noisy}(k,t)=\big[\bm{d}(k,t)+\bm{h}(t)\big]\cdot\bm{\sigma}$. First, we consider that $w(t)$ is a temporally fluc\-tua\-ting random variable sampled independently at each time step from a uniform distribution $w(t)\in[-W, W]$. We exa\-mi\-ne the scenarios $\bm{h}||\hat{\bm{z}}$ and $\bm{h}||\hat{\bm{x}}$. The former explicitly breaks CS due to the orientation of $\bm{h}$, while the latter only due to the noisy time profile. In Fig.~\ref{fig:Figure2} we illustrate the deviation $\sigma_{\rm LQP}-\nu_1$, which is expressed in units of $e^2/h$, after ave\-ra\-ging it over several random noise realizations. We find that this quantity is stable against noise strengths $W$ which are significantly smaller than $E_g\sim1$. For $\bm{h}||\hat{\bm{z}}$, the noise tends to affect the inference of this quantity more strongly than when $\bm{h}||\hat{\bm{x}}$. We also define a dimensionless measure that quantifies the arising CS violation (CSV) and reads as:
\bea
{\rm CSV}\equiv\frac{\hbar}{e^2}\lim_{A,T_2\rightarrow0}\frac{1}{T_2}\frac{\delta}{\delta A}\int_{-\pi}^{+\pi}\frac{dk}{2\pi}\,{\rm Re}\,\widetilde{\cal G}_{\rm CS}(k,\omega=0).\quad
\eea

\noi Note that ${\rm CSV}=0$ when CS is preserved. From Fig.~\ref{fig:Figure2} we infer that the CSV grows from zero upon increasing the noise strength, thus reflecting the breakdown of CS.

\begin{figure}[t!]
\centering
\includegraphics[width=\linewidth]{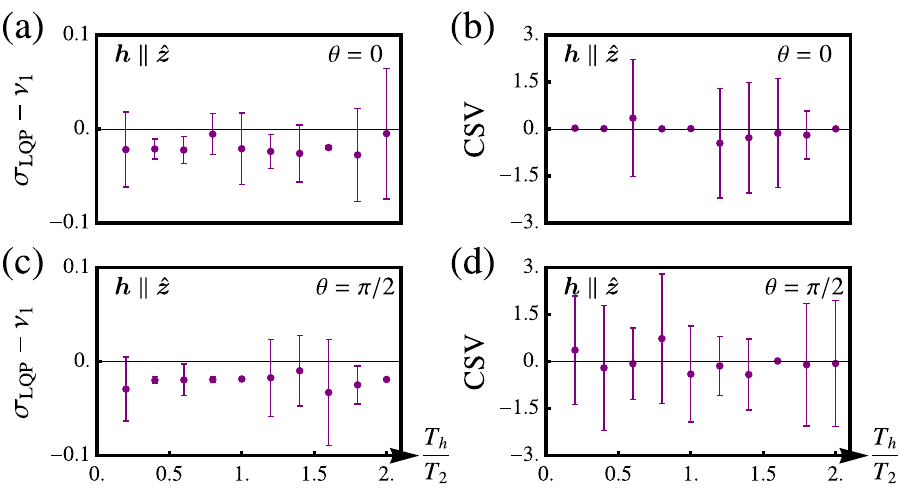} \\
\caption{Results for a harmonic noise profile where $w(t)=D\sin(2\pi t/T_{h}+\theta)$ with a random amplitude $D\in [-0.4,0.4]$. The model and remaining parameters are identical to those of Fig.~\ref{fig:Figure2}. We carried out numerics for $\theta=0$ and $\theta=\pi/2$ but found little difference between them. Panels (a) and (c) show the averaged deviation for $\sigma_{\rm LQP}-\nu_1$ while panels (b) and (d) present the respective CSV measures. We only show outcomes for $\bm{h}||\hat{\bm{z}}$, since for $\bm{h}||\hat{\bm{x}}$ the absolute values of the above quantities are $\sim0.01$. We mainly find that, when $\bm{h}||\hat{\bm{z}}$, the noise suppresses more drastically the quantization of $\sigma_{\rm LQP}$.}
\label{fig:Figure3}
\end{figure}

Apart from the above type of noise, we also consider the periodic profile $w(t)=D\sin\big(2\pi t/T_{h}+\theta\big)$ for $\theta=\{0,\pi/2\}$ and $\bm{h}||\{\hat{\bm{z}},\hat{\bm{x}}\}$, with $D$ being a random number. Here, although CS is expected to be preserved for the combinations $(\theta,\bm{h})=\{(0,||\hat{\bm{z}}),(\pi/2,||\hat{\bm{x}})\}$, results with sufficiently weak noise strength show remarkable stability for $\bm{h}||\hat{\bm{x}}$, independently on the value of $\theta$. In contrast, one finds that larger deviations appear in $\sigma_{\rm LQP}-\nu_1$ and CSV for $\bm{h}||\hat{\bm{z}}$, as it is shown in Fig.~\ref{fig:Figure3}.

\textit{\bt{Discussion} -} By introducing the concept of loop quenches, we unveil previously unknown scenarios for dynamical phases protected by chiral symmetry. Our work shows that although the Hamiltonian and the evolution operator generally belong to different symmetry classes, the symmetry-protected topology of the Hamiltonian can be yet recovered, thus, providing an exception to the known classifications of out-of-equilibrium SPT phases~\cite{McGinley2018,McGinley2019}. It yet remains an open question how the loop quench protocol affects the dynamics of the to\-po\-lo\-gi\-cal boundary modes of the initial Hamiltonian~\cite{Sedlmayr2018,Ghosh2023}, as well as the entanglement spectra~\cite{Gong2018,Chang2018,Pastori2020,Lane2024}.

Although several works have investigated the properties of quenches using
the return~\cite{DeLuca2014,Vajna2015,Nakagawa2016,Rossi2022quenchedSSH} or transition~\cite{Wong2022} rates, as well as other quantities specific to systems with chiral symmetry~\cite{Maffei2018,DErrico2020}, we are not aware of others employing the Loschmidt chirality amplitude $\widetilde{\cal G}_{\rm CS}(k,\omega)$ to diagnose
the topological properties of the target system. Our approach can be generalized for systems with additional bands, while our formulation can form the basis for studying similar pro\-per\-ties in higher dimensions. Although we focused on protocols which rely on the conservation of electric charge, our approach can be suitably extended to systems conser\-ving other U(1) symmetries, e.g., those which are invariant under spin rotations. We hope that our work will motivate theorists and experimentalists to employ loop-quench-probe protocols for the investigation of symmetry-protected phases.

\textit{\bt{Acknowledgments} -} We thank Kristian Patrick, Thomas Lane, and Hongqi Xu, for numerous inspiring and helpful discussions. In addition, N.~F. and M.~H. were supported by the Beijing Natural Science Foundation (Grant No. IS24022).

\textit{\bt{Data availability} -} The data that support the fin\-dings of this Letter are available from the authors upon reasonable request.


\begin{thebibliography}{99}

\bibitem{Hasan2010} M. Z. Hasan and C. L. Kane, \textit{Colloquium: Topological Insulators}, Rev. Mod. Phys. \bt{82}, 3045 (2010).

\bibitem{Qi2011} X.-L. Qi and S.-C. Zhang, \textit{Topological Insulators and Superconductors}, Rev. Mod. Phys. \bt{83}, 1057 (2011).

\bibitem{KongRev} L. Kong and H. Ding, \textit{Emergent vortex Majorana zero mode in iron-based superconductors}, Acta. Phys. Sin. \bt{69}, 110301 (2020).

\bibitem{AsbothBook} J. K. Asb\'oth, L. Oroszl\'any, and A. P\'alyi, \textit{A Short Course on Topological Insulators: Band-structure topology and edge states in one and two dimensions}, Lect. Notes Phys. \bt{919} (2016).

\bibitem{Senthil2015} T. Senthil, \textit{Symmetry-Protected Topological Phases of Quantum Matter}, Annu. Rev. Condens. Matter Phys. \bt{6}, 299 (2015).

\bibitem{Wen2017} X.-G. Wen, \textit{Colloquium: Zoo of quantum-topological phases of matter}, Rev. Mod. Phys. \bt{89}, 041004 (2017).

\bibitem{Ryu2010} S. Ryu, A. P. Schnyder, A. Furusaki, and A. W. Ludwig, \textit{Topological Insulators and Superconductors: Ten-Fold Way and Dimensional Hierarchy}, New J. Phys. \bt{12}, 065010 (2010).

\bibitem{Teo2010} J. C. Y. Teo and C. L. Kane, \textit{Topological Defects and Gapless Modes in Insulators and Superconductors}, Phys. Rev. B \bt{82}, 115120 (2010).

\bibitem{KitaevClassi} A. Kitaev, \textit{Periodic table for topological insulators and superconductors}, AIP Conf. Proc. \bt{1134}, 22 (2009).

\bibitem{Fidkowski} L. Fidkowski and A. Kitaev, \textit{Effects of interactions on the topological classification of free fermion
systems}, Phys. Rev. B \bt{81}, 134509 (2010).

\bibitem{Vajna2014} S. Vajna and B. D\'ora, \textit{Disentangling dynamical phase transitions from equilibrium phase transitions}, Phys. Rev. B \bt{89}, 161105(R) (2014).

\bibitem{DeLuca2014} A.~De Luca, \textit{Quenching the magnetic flux in a one-dimensional fermionic ring: Loschmidt echo and edge singularity}, Phys. Rev. B \bt{90}, 081403 (2014).

\bibitem{Vajna2015} S. Vajna and B. D\'ora, \textit{Topological classification of dynamical phase transitions}, Phys. Rev. B \bt{91}, 155127 (2015).

\bibitem{Nakagawa2016} Y. O. 
Nakagawa, G. Misguich, and M. Oshikawa, \textit{Flux quench in a system of interacting spinless fermions in one dimension}, Phys.
Rev. B \bt{93}, 174310 (2016).

\bibitem{Yang2018} C. Yang, L. Li, and S. Chen, \textit{Dynamical topological invariant after a quantum quench}, Phys. Rev. B \bt{97}, 060304 (2018).

\bibitem{Wang2017} C. Wang, P. Zhang, X. Chen, J. Yu, and H. Zhai, \textit{Scheme to Measure the Topological Number of a Chern Insulator from Quench Dynamics}, Phys.
Rev. Lett. \bt{118}, 185701 (2017).

\bibitem{Sun2018} W. Sun, C.-R. Yi, B.-Z. Wang, W.-W. Zhang, B. C.
Sanders, X.-T. Xu, Z.-Y. Wang, J. Schmiedmayer,
Y. Deng, X.-J. Liu, S. Chen, and J.-W. Pan, \textit{Uncover To\-po\-lo\-gy by Quantum Quench Dynamics}, Phys. Rev.
Lett. \bt{121}, 250403 (2018).


\bibitem{Hu2020} H. Hu and E. Zhao, \textit{Topological Invariants for Quantum Quench Dynamics from Unitary Evolution}, Phys. Rev. Lett. \bt{124}, 160402 (2020).

\bibitem{Zhang2020} L. Zhang, L. Zhang, and X.-J. Liu, \textit{Unified Theory to Characterize Floquet Topological Phases by Quench Dynamics}, Phys. Rev. Lett. \bt{125}, 183001 (2020).

\bibitem{Rossi2022quenchedSSH} L. Rossi and F. Dolcini, \textit{Nonlinear current and dynamical quantum phase transitions in the flux-quenched Su-Schrieffer-Heeger model}, Phys. Rev. B \bt{106}, 045410 (2022).

\bibitem{Qiu2024} L. Qiu, L.-K. Lim, and X. Wan, \textit{Dynamical geometry of the Haldane model under a quantum quench}, Phys. Rev. B \bt{110},
104311 (2024).


\bibitem{McGinley2018} M. McGinley and N. R. Cooper, \textit{Topology of One-Dimensional Quantum Systems Out of Equilibrium}, Phys. Rev. Lett. \bt{121}, 090401 (2018).

\bibitem{Gong2018} Z. Gong and M. Ueda, \textit{Topological Entanglement-Spectrum Crossing in Quench Dynamics}, Phys. Rev. Lett. \bt{121}, 250601 (2018).

\bibitem{Sedlmayr2018} N. Sedlmayr, P. Jaeger, M. Maiti and J. Sirker, \textit{Bulk-boundary correspondence for dynamical phase transitions in one-dimensional topological insulators and superconductors},
Phys. Rev. B \bt{97}, 064304 (2018).

\bibitem{McGinley2019} M. McGinley and N. R. Cooper, \textit{Classification of topological insulators and superconductors out of equilibrium}, Phys. Rev. B \bt{99}, 075148 (2019).

\bibitem{Pastori2020} L. Pastori, S. Barbarino, and J. C. Budich, \textit{Signatures of topology in quantum quench dynamics and their interrelation},
Phys. Rev. Res. \bt{2}, 033259 (2020).

\bibitem{Ghosh2023} A. Ghosh, A. M. Martin and S. Majumder, \textit{Quench dynamics of edge states in a finite extended Su-Schrieffer-Heeger system}, Phys. Rev. E \textbf{108}, 034102 (2023).

\bibitem{Lane2024} T.~L.~M.~Lane, M Horv\'ath, and K. Patrick, \textit{Extended edge modes and disorder preservation of a symmetry-protected topological phase out of equilibrium}, Phys. Rev. B \bt{110}, 165139 (2024).

\bibitem{Asboth2013} J. K. Asb\'oth and H. Obuse, \textit{Bulk-boundary correspondence for chiral symmetric quantum walks}, Phys. Rev. B \bt{88}, 121406(R) (2013).

\bibitem{Asboth2014} J. K. Asb\'oth, B. Tarasinski, and P. Delplace, \textit{Chiral symmetry and bulk-boundary correspondence in pe\-rio\-di\-cal\-ly driven one-dimensional systems}, Phys. Rev. B \bt{90}, 125143 (2014).

\bibitem{Nathan2015}  F. Nathan and M.~S. Rudner, \textit{Topological singularities and the general
classification of Floquet–Bloch systems}, New J. Phys. \bt{17} 125014 (2015).

\bibitem{Fruchart2016} M. Fruchart, \textit{Complex classes of periodically driven topological lattice systems}, Phys. Rev. B \bt{93} 115429 (2016).

\bibitem{RoyHarper1} R. Roy and F. Harper, \textit{Floquet topological phases with symmetry in all dimensions}, Phys. Rev. B \bt{95}, 195128 (2017).

\bibitem{RoyHarper2} R. Roy and F. Harper, \textit{Periodic table for Floquet topological insulators}, Phys. Rev. B \bt{96}, 155118 (2017).

\bibitem{Yao2017} S. Yao, Z. Yan, and Z. Wang, \textit{Topological invariants of Floquet systems: General formulation, special properties, and Floquet topological defects}, Phys. Rev. B \bt{96}, 195303 (2017).

\bibitem{Kennes2019} D. M. Kennes, N. M\"uller, M. Pletyukhov, C. Weber, C. Bruder, F. Hassler, J. Klinovaja, D. Loss, and H. Schoeller, \textit{Chiral one-dimensional Floquet topological insulators beyond the rotating wave approximation}, Phys. Rev. B \bt{100}, 041103(R) (2019).

\bibitem{Assili2024} M. Assili and P. Kotetes, \textit{Dynamical Chiral Symmetry and Symmetry-Class Conversion in Floquet Topological Insulators}, Phys. Rev. B \bt{109}, 184307 (2024).

\bibitem{Cardoso2025} G. Cardoso, H.-C. Yeh, L. Korneev, A. G. Abanov, and A. Mitra, Phys. Rev. B \bt{111}, 125162 (2025).


\bibitem{OkaRev} T. Oka and S. Kitamura, \textit{Floquet Engineering of Quantum Materials}, Annu. Rev. Condens. Matter Phys. \bt{10}, 387 (2019).

\bibitem{RudnerRev} M. S. Rudner and N. H. Lindner, \textit{Band structure engineering and non-equilibrium dynamics in Floquet topological insulators}, Nat. Rev. Phys. \bt{2}, 229 (2020).

\bibitem{NiuRev} D. Xiao, M.-C. Chang, and Q. Niu, \textit{Berry phase effects on electronic properties}, Rev. Mod. Phys. \bt{82}, 1959 (2010).

\bibitem{Chang2018} P.-Y. Chang, \textit{Topology and entanglement in quench dynamics}, Phys. Rev. B \bt{97}, 224304 (2018).

\bibitem{Wong2022} C.-Y. Wong and W.-C. Yu, \textit{Loschmidt amplitude spectrum in dynamical quantum phase transitions}, Phys. Rev. B \bt{105}, 174307 (2022).

\bibitem{Maffei2018}
M. Maffei, A. Dauphin, F. Cardano, M. Lewenstein, and
P. Massignan, \textit{Topological characterization of chiral mo\-dels
through their long time dynamics}, New J. Phys. \bt{20}, 013023 (2018).

\bibitem{DErrico2020} A. D’Errico \textit{et al.}, \textit{Bulk detection of time-dependent topological transitions in quenched chiral models}, Phys. Rev. Res. \bt{2}, 023119 (2020).

\end{thebibliography}
\end{document}